\documentclass[preprint,letterpaper,amsmath,amssymb,aps,prb,showpacs]{revtex4-1}
\usepackage{graphicx}
\usepackage{amsmath}
\usepackage{amsfonts}
\usepackage{amssymb}
\usepackage{hyperref}
\usepackage{natbib}
\begin{document}
\title{An Explicit Method for Schrieffer-Wolff Transformation}
\author{Rukhsan Ul Haq }
\affiliation{Theoretical Sciences Unit - Jawaharlal Nehru Center for Advanced Scientific Research}
\author{ Sachin Satish Bharadwaj}
\email{satish@jncasr.ac.in}
\affiliation{Theoretical Sciences Unit - Jawaharlal Nehru Center for Advanced Scientific Research\\ National Centre for Biological Sciences - TIFR\\ Centre for Fundamental Research and Creative Education\\ Department of Mechanical Engineering R V College of Engineering\\ Bangalore,India}
\author{and\\Towseef Ali Wani}
\affiliation{International Centre for Theoretical Sciences-TIFR, India}
\begin{abstract}
   Schrieffer-Wolff transformation is a very important transformation in Quantum Many Body physics. Yet, there isn't an explicit method in the literature to calculate the generator of this unitary transformation directly from the hamiltonian. In this paper we present an explicit method to compute the generator of the Schrieffer-Wolff transformation in general and then calculate the same for two very  important hamiltonians in condensed matter physics, the Single Impurity Anderson model (SIAM)  and its lattice generalization , the Periodic Anderson Model (PAM). Our method can be used to carry out the transformation directly from the hamiltonian and the generator of the transformation can be obtained without any recourse to perturbative expansion or intuitive guesswork. In fact here, the full generator is calculated rather than its expansion to first two orders which is usually the case.
\end{abstract}
\maketitle
\section{Introduction} 
Schrieffer-Wolff transformation(SW) was actually introduced in\cite{SW} to relate Anderson impurity model and the Kondo model. Since then, this transformation has been used very extensively in Condensed matter physics. One could see\cite{Bravyi} as a recent reference in the literature on SW transformation where the authors have given a mathematically rigorous treatment of the transformation. SW transformation is a unitary transformation which removes the off-diagonal terms to the first order and hence serves as a way of diagonalization. Though this transformation is routinely used by researchers working in the condensed matter community, it was surprising for us to find that there is no explicit method to find the generator of this transformation. In this paper we present an explicit method to get the generator of this transformation and we aslo calculate the same for two very important hamiltonians of condensed matter physics, namely the Single Impurity Anderson model(SIAM) for which the transformation was carried out in the original paper\cite{SW}  and its lattice generalization , the Periodic Anderson Model(PAM). The rest of the paper is organized as follows: In the  section II we give a brief introduction to Schrieffer-Wolff transformation and how in literature, there are different approaches and interpretations for the same. In section III we describe the method to obtain the generator and then in section IV and V we calculate the generators for the said Hamiltonians. We conclude by summarizing the results and discussions.

\section{Schrieffer-Wolff Transformation : Formalism}
Unitary transformation is the standard method for diagonalization in Quantum Mechanics and Condensed matter physics\cite{Wagner}. By the unitary transformation one changes to 
the basis in which the given hamiltonian becomes diagonal and one achieves diagonalization in a one step process. However, this is not possible for all hamiltonains. So, in latter case, one tries to diagonalize the hamiltonian in a perturbative manner. One can still use unitary transformations to achieve this goal and Schrieffer-Wolff transformation is such a method. It has been used extensively in different areas of physics under different names\cite{Bravyi}. In relativistic Quantum Mechanics  it is called Foldy-Wutheysen transformation\cite{FW},in Semiconductor physics it is called k.p perturbation theory\cite{Winkler} and in condensed matter physics it has been used as Frohlich transformation for electron-phonon problem\cite{Frohlich}.
Schrieffer-Wolff transformation not only diagonalizes the hamiltonian in a perturbative manner in which case it is unitary perturbation theory but it also renormalizes the 
parameters in the hamiltonian and hence can be thought of as a kind of renormalization procedure. SW transformation in the latter sense is used to get the effective hamiltonian of the given hamiltonian and hence it takes us to a particular regime of a hamiltonain in its parameter space and it is in the same sense that SW transformation was used in\cite{SW}.\\
Schrieffer-Wolff transformation is a unitary transformation. So one chooses the proper unitary operator which can either fully diagonalize the hamiltonian or to some desired order.  
\begin{align}
H'&=U^{\dag}HU \\
H'& = e^{S}H e^{-S}
\end{align}
where S is the generator of this transformation and is an anti-hermitian operator. Usually one requires of this transformation to cancel the off-diagonal terms to the first order so that following condition is satisfied.
\begin{equation}
\left[S,H_{0}\right]=-H_{v}
\end{equation}
Expanding the operator exponential using BCH formula one gets series expansion for the transformed hamiltonian $H' $
\begin{equation}
H'=H_{0}+\frac{1}{2}\left[S,H_{v}\right]+\frac{1}{3}\left[S,\left[S,H_{v}\right]\right]+....
\end{equation}
where$H_{0}$ and $H_{v}$ are diagonal and off-diagonal parts of the hamiltonian $H$. Since the off-diagonal term gets cancelled to the first order so the effective hamiltonian to the second order is given by
\begin{equation}
H_{eff} = H_{0} + \frac{1}{2}\left [S,H_{v} \right]
\end{equation}
 
Schrieffer-Wolff transformation being very important transformation has been generalized in various ways. One very important generalization was done by Wegner\cite{Wegner} and Glazek and Wilson\cite{Glazek} independently. The new method has been called Flow equation method by Wegner and Similarity Renormalization by Glazek and Wilson. In flow equation method the unitary transformation is once again used but it is done in a continuous fashion because the generator depends on the flow parameter. The relation between SW transformation and flow equation method has been worked out in \cite{Kehrein}. SW transformation has  been generalized to dissipative quantum systems as well.\cite{Kessler} \\
In  addition to the method of SW transformation we have described above, where we have used a generator to carry out the transformation, there is yet another method of doing the transformation in which one uses Projection operators\cite{Hewson} or Hubbard operators\cite{Fazekas}. Though with the projection operator method we can obtain the effective hamiltonian, 
it does not recover all the terms which one gets in the generator method. Projection operator method is in spirit of SW transformation being a renormalization procedure, while the 
Hubbard operator method is particularly suited for Hubbard model.
\vspace{-0.2cm}
\section{Calculation of the Generator for the Schrieffer-Wolff Transformation  }
The most crucial step in doing the SW transformation is to get the generator of the transformation. Once the generator is calculated the rest of the calculation is quite 
straightforward. So having an explicit method for calculating the generator is of immense value. Here, we would emphasize that in the literature there is no explicit method to 
calculate the generator and hence to carry out the transformation directly from the hamiltonian. In \cite{Wagner} \cite{Coleman} generator itself is obtained in a perturbative 
manner. So, one does not obtain the generator, rather one develops it in an orderly fashion. In \cite{Phillips} the generator is obtained by guessing it, which assumes some 
experience within such a method. In \cite{Gulasci} though the authors have summed the series to all orders, they have not described how they compute the generator in the first place. \\
In this section, we  present the method for a general hamiltonian and then in the following sections we will calculate the generator of specific Hamiltonians using this explicit method.\\
  Let H be our full hamiltonian and $H_{0}$ be the diagonal part and $H_{v}$ be the off-diagonal part of the full  hamiltonian. To obtain the generator we will proceed in two 
steps. In the first step, we will find the commutator $[H_{0},H_{v}]$ and call it $\eta$. In the second step we will impose the condition of removing the off-diagonal part up till the
 first order on $\eta$. To do that, we will have to keep the coefficients undetermined and they will be eventually determined by the above condition. So $\eta$ has to satisfy $[\eta,H_{0}]= -H_{v}$ in order to be the generator of the transformation. The latter condition determines the coefficients and we finally get the generator for SW transformation for a given 
hamiltonian. In the next section, we will calculate the generator of SW transformation for the Single Impurity  Anderson Model for which the transformation was carried out in the 
original paper\cite{SW}.
\section{Generator for Anderson Impurity Model}
 We will first write down the Single Impurity Anderson Model hamitonian in second quantized notation:
\begin{equation} 
H = \sum_{k\sigma}\epsilon_{k}c^{\dag}_{k\sigma}c_{k\sigma}+\sum_{\sigma}\epsilon_{d}d^{\dag}_{\sigma}d_{\sigma}+\sum_{k\sigma}V_{k}(c^{\dag}_{k\sigma}d_{\sigma}+d^{\dag}_{\sigma}c_{k\sigma})+ Un_{d\uparrow}n_{d{\downarrow}}
\end{equation}
The hamiltonian has one off-diagonal term which we call as $H_{v}$ and diagonal terms which together we call $H_{0}$.
\begin{equation}
H_{0}=\sum_{k\sigma}\epsilon_{k}c^{\dag}_{k\sigma}c_{k\sigma}+\sum_{\sigma}\epsilon_{d}d^{\dag}_{\sigma}d_{\sigma}+ Un_{d\uparrow}n_{d{\downarrow}} 
\end{equation}
\begin{equation}
H_{v}=\sum_{k\sigma}V_{k}(c^{\dag}_{k\sigma}d_{\sigma}+d^{\dag}_{\sigma}c_{k\sigma})
\end{equation}
Now the first step is to calculate $\eta$ which is basically commutator of diagonal part with off-diagonal part of the hamiltonian.
\begin{align}
\eta & = \left [H_{0},H_{v}\right ] \\
\eta & =\left [\sum_{k\sigma}\epsilon_{k}c^{\dag}_{k\sigma}c_{k\sigma}+\sum_{\sigma}\epsilon_{d}d^{\dag}_{\sigma}d_{\sigma}+
Un_{d\uparrow}n_{d{\downarrow}},\sum_{k\sigma}V_{k}(c^{\dag}_{k\sigma}d_{\sigma}+d^{\dag}_{\sigma}c_{k\sigma})\right] \\
\eta & = \sum_{k\sigma}(\epsilon_{k}-\epsilon_{d}-Un_{d\bar{\sigma}})V_{k}(c^{\dag}_{k\sigma}d_{\sigma}-d^{\dag}_{\sigma}c_{k\sigma})
\end{align}
In the second step we will impose the condition of removing the off-diagonal term to the first order. To do that, we will keep the
coefficients undetermined and actually they will get determined automatically once $\eta$ satisfies the condition. We will label it with S  to emphasize that it is not actually $\eta$ which is the generator rather it is S with correct coefficients. What $\eta$ has similar to S is the form of the operators.
\begin{equation}
S =\sum_{k\sigma}(A_{k}-B_{k}n_{d\bar{\sigma}})V_{k}(c^{\dag}_{k\sigma}d_{\sigma}-d^{\dag}_{\sigma}c_{k\sigma})
\end{equation}
Now we will impose the condition on S to determine $A_{k}$ and $ B_{k}$
\begin{align}
&\left [S,H_{0}\right ]   =  - H_{v} \\
&\Rightarrow \left [ \sum_{k\sigma} A_{k}(\epsilon_{d}-\epsilon_{k}) +
  \sum_{k\sigma}(A_{k}U-B_{k}(\epsilon_{d}-\epsilon_{k}+U)n_{d\bar{\sigma}})\right ]  
  (V_{k}(c^{\dag}_{k\sigma}d_{\sigma}+  d^{\dag}_{\sigma}c_{k\sigma}) \\
& =  -\sum_{k\sigma} V_{k}(c^{\dag}_{k\sigma}d_{\sigma} + d_{\sigma}c^{\dag}_{k\sigma}) \\
&\Rightarrow A_{k}(\epsilon_{d}-\epsilon_{k}) + (A_{k}U+ B_{k}(\epsilon_{d}-\epsilon_{k}+U)n_{d\bar{\sigma}} =  -1 
\end{align}
Solving for $A_{k}$ and $B_{k}$ we obtain:
\begin{align}
A_{k}& =\frac{1}{\epsilon_{k}-\epsilon_{d}}\\
B_{k}& = \frac{1}{\epsilon_{k}-\epsilon_{d}-U} - \frac{1}{\epsilon_{k}-\epsilon_{d}}
\end{align}
In this way we have calculated the generator of SW transformation for Single Impurity Anderson Model.
\begin{equation}
S = \sum_{k\sigma}(A_{k} + B_{k}n_{d\bar{\sigma}})V_{k}(c^{\dag}_{k\sigma}d_{\sigma}-d^{\dag}_{\sigma}c_{k\sigma}) 
\end{equation}
\begin{align}
A_{k} & = \frac{1}{\epsilon_{k}-\epsilon_{d}} \\ 
B_{k} & = \frac{1}{\epsilon_{k}-\epsilon_{d}-U} - \frac{1}{\epsilon_{k}-\epsilon_{d}}
\end{align}
We can write the generator in the same form as it was written in\cite{SW} by using extra index $\alpha$ which takes two values.
\begin{equation}
\mathrm{S} =\sum_{k\sigma\alpha}\frac{V_{k}}{\epsilon_{k}-\epsilon_{\alpha}}n^{\alpha}_{d,\bar{\sigma}}c^{\dag}_{k\sigma}d_{\sigma}-h.c.
\end{equation} 
\begin{align}
n_{d\bar{\sigma}}^{\alpha} & = n_{d\bar{\sigma}} & \epsilon_{\alpha}=\epsilon_{d}+U \quad \alpha = + \\
&=1-n_{d\bar{\sigma}}    &\epsilon_{\alpha}=\epsilon_{d}      \quad  \alpha = -
\end{align}
Summing over $\alpha$ we get $\mathrm{S}$ in the same form as we have calculated.
\begin{align}
\mathrm{S} & = \sum_{k\sigma}[\frac{V_{k}}{\epsilon_{k}-\epsilon_{d}-U} n_{d\bar{\sigma}}c^{\dag}_{k\sigma}d_{\sigma}+ \\
& \frac{V_{k}}{\epsilon_{k}-\epsilon_{d}}(1-n_{d\bar{\sigma}})c^{\dag}_{k\sigma}d_{\sigma}] - h.c.
\end{align}
Rearranging the terms one gets $\mathrm{S}$ in the form as we have calculated above. Further, the detailed SW Transformation for SIAM could be found in the paper. \cite{Sachin}
\section{Generator for Periodic Anderson Model}
Periodic Anderson Model(PAM) is the standard model in heavy fermion physics. We will write down the model in standard second quantized notation as given in \cite{Nolting}
\begin{equation}
H = \sum_{k\sigma}\epsilon_{k}n_{k\sigma}+\sum_{i\sigma}\epsilon_{f}n^{f}_{i\sigma}+U\sum_{i}n_{i\uparrow}n_{i\downarrow}+
\sum_{ki\sigma}(V_{k}e^{-ikR_{i}}c^{\dag}_{k\sigma}f_{i\sigma}+h.c)
\end{equation}
To calculate the generator of SW transformation for PAM we will proceed as per the method given in section III. So we will first calculate $\eta$ as follows:
\begin{align}
\eta & = [H_{0},H_{v}] \\
&\eta =\left [\sum_{k\sigma}\epsilon_{k}n_{k\sigma}+\sum_{i\sigma}\epsilon_{f}n^{f}_{i\sigma}+U\sum_{i}n_{i\uparrow}n_{i\downarrow},\sum_{ki\sigma}(V_{k}e^{-ikR_{i}}c^{\dag}_{k\sigma}f_{i\sigma}+h.c)\right ] \\
&\eta = \sum_{ki\sigma}(\epsilon_{k}-\epsilon_{f}-Un_{i\bar{\sigma}})V_{k}(c^{\dag}_{k\sigma}f_{i\sigma}e^{-ikR_{i}} - h.c.)
\end{align}
So we can write the generator for PAM as:
\begin{equation}
S = \sum_{ki\sigma}(A_{k} + B_{k}n_{i\bar{\sigma}})V_{k}(c^{\dag}_{k\sigma}f_{i\sigma}e^{-ikR_{i}}-h.c)
\end{equation}
Where $A_{k}$ and $B_{k}$ need to be determined. To do so we will follow the second step of the method and proceed as follows.
\begin{align}
&\left[S,H_{0}\right] =  - H_{v} \\
 &\left[\sum_{ki\sigma}(A_{k}+ B_{k}n_{i\bar{\sigma}})V_{k}(c^{\dag}_{k\sigma}f_{i\sigma}e^{-ikR_{i}}-h.c),  \sum_{k'\sigma'}\epsilon_{k'}n_{k'\sigma'}+
\sum_{i\sigma'}\epsilon_{f} n_{i\sigma'}+U\sum_{i}n_{i\uparrow}n_{i\downarrow} \right] \\
& = - H_{v}\\ 
&\Rightarrow \sum_{ki\sigma}(A_{k}(\epsilon_{f}-\epsilon_{k}+Un_{i\bar{\sigma}})+B_{k}n_{i\bar{\sigma}}(\epsilon_{f}-\epsilon_{k}+U)
V_{k}(c^{\dag}_{k\sigma}f_{i\sigma}e^{-ikR_{i}} + h.c.))\\
&= - V_{K}  (c^{\dag}_{k\sigma}f_{i\sigma}e^{-ikR_{i}} +h.c.)\\
&\Rightarrow A_{k}(\epsilon_{f}-\epsilon_{k})+( A_{k}U +B_{k}(\epsilon_{f}-\epsilon_{k}+U))n_{i\bar{\sigma}}= -1 
\end{align}
Solving for $A_{k}$ and $B_{k}$ in the above equation we get:
\begin{align}
 A_{k}& = \frac{1}{\epsilon_{k}-\epsilon_{f}} \\
 B_{k}& = \frac{1}{\epsilon_{k}-\epsilon_{f}-U} - \frac{1}{\epsilon_{k}-\epsilon_{f}}
\end{align}
This way we have got the generator of SW transformation for PAM and can be written as : 
\begin{equation}
S= \sum_{k\sigma i}(A_{k}+B_{k}n_{i\bar{\sigma}})V_{k}(c^{\dag}_{k\sigma}f_{i\sigma}e^{-ikR_{i}}-f^{\dag}_{i\sigma}c_{k\sigma}e^{ikR_{i}})
\end{equation}
where $A_{k}$ and $B_{k}$ are given by 
\begin{align}
A_{k} & = \frac{1}{\epsilon_{k}-\epsilon_{f}}\\
B_{k} & = \frac{1}{\epsilon_{k}-\epsilon_{f}-U}-\frac{1}{\epsilon_{k}-\epsilon_{f}}
\end{align}
Further, the detailed SW Transformation for SIAM could be found in the paper.\cite{Sachin}

\section{Summary}
Schrieffer-Wolff(SW) transformation is a very important transformation in quantum many-body physics, where it is routinely used to get the low energy effective hamiltonian of quantum many-body hamiltonians which are in general difficult to be dealt analytically. Historically SW transformation has given a very important insight into the understanding of the low energy sector of Anderson Impurity models which needs advanced methods like conformal field theory for its solution. Similarly, SW transformation was also used to understand the electron-phonon problem and how it gives rise to attractive effective interaction in BCS superconductivity. Given the significance of SW transformation it is surprising that there is no explicit method to calculate the generator of this transformation directly from the hamiltonian without any recourse to perturbative expansion or intuitive guesswork. In this paper, we have presented such an  explicit method and have demonstrated this new method, by carrying out the SW transformation of Single Impurity Anderson Model and its lattice version, the Periodic Anderson Model.

\begin{acknowledgements}
The authors would like to acknowledge Professor N S Vidhyadhiraja, TSU , JNCASR for his valuable discussions, guidance and insights into this work and also would like to thank the Department of Science and Technology(DST), India for the funding and JNCASR for providing the very conducive environment to carry out this research in the institute.
\end{acknowledgements}


\end{document}